# *In-situ* angle-resolved photoemission study of MBE-grown $(La, Ce)_2CuO_4$ thin films

H. Yamamoto*, M. Naito, A. Tsukada, S. Suzuki

NTT Basic Research Laboratories, NTT Corporation, 3-1 Wakamiya Morinosato, Atsugi-shi, Kanagawa 243-0198, Japan

Abstract

We have performed *in-situ* angle-resolved photoemission spectroscopy on the electron doped cuprate superconductor $La_{2-x}Ce_xCuO_4$ (x = 0.1) using high-quality films grown by molecular beam epitaxy.  The obtained photoemission spectra were free from "dirt" peaks.  In addition, the time evolution of the spectra due to impurity adsorption is rather gradual, and the sample surface is stable even at ambient temperature, which is in contrast with previous reports on bulk $Nd_{2-x}Ce_xCuO_4$ (NCCO) single crystals.  The energy distribution curves show clear dispersions near the Fermi energy ($E_F$) with some $E_F$ crossings.  The resultant Fermi surface is hole-like and centered around $(\pi, \pi)$, which essentially agrees with that predicted for NCCO by band calculations and also with the experimentally determined Fermi surface of optimally doped NCCO (x = 0.15).

*Corresponding author.

Dr. Hideki Yamamoto

Postal address: Superconducting Thin Films Research Group, NTT Basic Research Labs., 3-1 Wakamiya Morinosato, Atsugi-shi, Kanagawa 243-0198, Japan

Phone: +81-46-240-3523

Fax: +81-46-240-4717

E-mail address: hideki@will.brl.ntt.co.jp

**1. Introduction**

Angle-resolved photoemission spectroscopy (ARPES) is useful for investigating the electronic structure of the high-$T_c$ superconductors and has provided a number of important results, such as observations of a rather conventional Fermi surface [1], a momentum-resolved superconducting gap [2], a pseudogap [3]. In fact, several review articles about ARPES on the high-$T_c$ superconductors have already been published [4 - 6]. However, the majority of the work has been performed on cleaved Bi-22(n-1)n single crystals and there is not sufficient data of equal quality on numerous other relevant compounds. This is mostly because of the difficulty in preparing a bulk-representative surface [7], which is crucial for ARPES due to its extremely short probing depth [8]. *In-situ* ARPES with thin film samples has the potential to overcome this problem. In addition, this approach also provides unique opportunities lacking in bulk ARPES; namely, ARPES on (1) flat and large area surfaces, (2) high-pressure phases such as infinite-layer compounds [9], (3) epitaxially strained samples [10], and (4) ultra-thin films with comparable thickness to the probing depth [11]. Very recently, Abrecht *et al*. have actually achieved thin-film ARPES using pulsed-laser-ablation films [12]. Their experiment revealed that epitaxially strained (La, Sr)$_2$CuO$_4$ films with a higher $T_c$ than "bulk $T_c$" have a different low-energy electronic structure from that of (unstrained) single crystals, which demonstrated that thin-film ARPES is one of the promising routes to clarify the electronic structure essential for high-$T_c$ superconductivity.

In this study, as the first step, we performed ARPES measurements of T'-La$_{2-x}$Ce$_x$CuO$_4$ (LCCO) films (x = 0.1, $T_c^{end}$ = 28 K) grown by molecular beam epitaxy (MBE). This compound is an *n*-type superconductor whose single crystals cannot be obtained by bulk growth [13]. In addition, LCCO has following unique characteristics [14]: the optimum $T_c^{end}$ is over 30 K, highest in the T' family; the optimum doping level is x ~ 0.09, which is significantly lower than x = 0.15 for Nd$_{2-x}$Ce$_x$CuO$_4$ (NCCO); the superconducting window is extended to a lower doping region than in NCCO. The comparison by ARPES of this singular doping dependence in LCCO with that reported for NCCO [15, 16] will be quite interesting.

## 2. Experimental

We grew LCCO films on SrTiO$_3$(001) substrates in a customer-designed MBE chamber from metal sources using multiple electron-gun evaporators with accurate stoichiometry control of the atomic beam fluxes [17]. After the growth, the films were reduced in vacuum to remove interstitial apical oxygen. Details of the sample preparations were described elsewhere [14]. For *in-situ* ARPES, the grown films were transferred to an analysis chamber using a customer-designed portable vacuum vessel (~ 10$^{-8}$ Torr) equipped with a non-evaporable getter pump, which enables vacuum transfer without any continuous power supply. The analysis chamber is located at a beamline of ABL-6B at the Normal-conducting Accelerating Ring of the NTT SOR [18]. An Mg filter was installed on the beamline to eliminate the higher-order diffracted light [19], which was actually useful for photon energies ranging from 25 to 50 eV. In the present study, all the ARPES measurements were performed with photon energy of 50 eV at ambient temperature using an SES-2002 electron spectrometer.

## 3. Results and discussion

Fig. 1 shows partially angle-integrated energy distribution curve (EDC) of the as-transferred specimens. Two features are noteworthy: the absence of a "dirt" peak around 9 eV [20], and the presence of a clear Fermi edge, both indicating that the sample surface prepared in our method is free from contamination and suitable for ARPES investigation of the low-energy electronic structure. Our preliminary experiments on $h\nu$ dependence (25 – 50 eV) indicated that the structure near Fermi energy ($E_F$) is prominent for ~50 eV photons, so the following ARPES data were obtained using 50 eV photons.

Next, we examined the time evolution of the spectra in ultra-high vacuum at ambient temperature. As shown in Fig. 2, components above ~5 eV BE gradually grow with time. However, the near-$E_F$ spectral weight is mostly preserved even after a week; that is, the surface of the LCCO film degrades gradually at ambient temperature. This is in contrast with a previous report on bulk NCCO single crystals, where the photoemission spectra had to be obtained at 10 K to avoid the considerable surface degradation that occurs at higher temperature [21]. This discrepancy may arise from differences in the surface structure and/or chemistry between our MBE-grown films and the cleaved single crystals.

Fig. 3 shows near-$E_F$ EDCs from high-symmetry lines in the Brillouin zone [(a) – (c)] together with one from a rather low symmetry line [(d)]. In Fig. 3(a), along the Γ to (π, π) direction, a broad feature disperses towards $E_F$, passing it at $k_F$ (0.47π, 0.47π), and then disappears. Similar dispersive features and Fermi surface crossings are also observed in Figs. 3(b) and (d). Note that the quasi-particle peaks, which should be sharp near $E_F$, are rather broad even around $k_F$ due most likely to the thermal broadening effect; *i.e.*, spectra were obtained at ambient temperature. Along the Γ to (π, 0) direction [Fig. 3(c)], only a small dispersive feature is discernible well below $E_F$ (~ 400 meV). The flat band feature reported for optimally doped *p*-type compounds [22] is not observed.

More comprehensive spectra taken over a large region in the momentum space enabled us to map an experimental FS as shown in Fig. 4. The results indicate LCCO (x = 0.1) has a hole-like Fermi surface centered at the (π, π) point, which essentially agrees with a Fermi surface for NCCO predicted by LDA band calculation [23] and also with the previous reports on optimally doped NCCO (x = 0.15) [15, 21, 24]. However, the observed Fermi surface is in contrast to that reported for NCCO with the same doping level (x = 0.1) [but at different temperature (~10 K)], where no EDCs cross $E_F$ except at around (π, 0) and its equivalent points. This is reasonable since LCCO is superconducting at x = 0.1, whereas NCCO is non-superconducting with antiferromagnetism persisting. Nevertheless, the apparent discrepancy between two materials suggests that more comprehensive ARPES studies on doping- and temperature-dependence for LCCO are imperative for a better understanding of the doping-induced "insulator-metal" crossover in the electron-doped cuprates. Such an attempt is currently under way.

## 4. Summary

We have performed ARPES experiments on $La_{2-x}Ce_xCuO_4$ ($x = 0.1$) using MBE-grown films. The photoemission spectra obtained *in situ* are free from "dirt" peaks and their degradation with time is rather gradual at ambient temperature. The ARPE spectra show clear dispersions with some $E_F$ crossings. The resultant Fermi surface essentially agrees with that predicted by LDA calculation and also with the experimentally determined Fermi surface of optimally doped $Nd_{2-x}Ce_xCuO_4$ ($x = 0.15$), in that it is hole-like and centered around ($\pi$, $\pi$). The combination of ARPES and MBE films may provide one of the best opportunities for studying the intrinsic electronic structure of a wide variety of high-$T_c$ superconductors.

**Acknowledgements**

The authors are grateful to Prof. S. Suga and Dr. A. Sekiyama of Osaka Univ., Dr. Y. Aiura and Dr. H. Eisaki of AIST, Prof. T. Takahashi and Prof. S. Suzuki of Tohoku Univ., Dr. T. Yokoya of Univ. of Tokyo, Dr. A. Ino of Hiroshima Univ., Prof. Z. –X. Shen of Stanford Univ., Prof. C. Kim of Yonsei Univ., and Dr. A. Matsuda of NTT for their kind advice on designing the ARPES system and stimulating discussions. They also thank Dr. K. Yamada, Dr. Y. Watanabe, Dr. F. Maeda, and Dr. T. Kiyokura of NTT for their contributions in the beamline construction and SOR operation, and Dr. S. Ishihara, Dr. H. Takayanagi, and Dr. M. Morita for their support and encouragement.

**References**

[1] T. Takahashi, H. Matsuyama, H. Katayama-Yoshida, Y. Okabe, S. Hosoya, K. Seki, H. Fujimoto, M. Sato, H. Inokuchi, Nature 334 (1988) 691.

[2] Z. –X. Shen, D. S. Dessau, B. O. Wells, D. M. King, W. E. Spicer, A. J. Arko, D. Marshell, L. W. Lombardo, A. Kapitulnik, P. Dickinson, S. Doniach, J. DiCarlo, A. G. Loeser, C. H. Park, Phys. Rev. Lett. 70 (1993) 1553.

[3] H. Ding, T. Yokoya, J. C. Campuzano, T. Takahashi, M. Randeria, M. R. Norman, T. Mochiku, K. Kadowaki, J. Giapintzakis, Nature 382 (1996) 51.

[4] Z. –X. Shen, D. S. Dessau, Phys. Rep. 253 (1995) 1.

[5] D. W. Lynch, C. G. Olson, Photoemission Studies of High-Temperature Superconductors, Cambridge University Press, Cambridge, 1999.

[6] A. Damascelli, Z. Hussain, Z. –X. Shen, Rev. Mod. Phys. 75 (2003) 473.

[7] H. Yamamoto, M. Naito, H. Sato, Phys. Rev. B 56 (1997) 2852.

[8] S. Hüfner, Photoelectron Spectroscopy, Springer-Verlag, Berlin, 1996.

[9] S. Karimoto, K. Ueda, M. Naito, T. Imai, Appl. Phys. Lett. 79 (2001) 2767.

[10] H. Sato, M. Naito, Physica C 274 (1997) 221.

[11] H. Sato, H. Yamamoto, M. Naito, Physica C 274 (1997) 227.

[12] M Abrecht, D. Ariosa, D. Cloetta, S. Mitrovic, M. Onellion, X. X. Xi, G. Margaritondo, D. Pavuna, Phys. Rev. Lett. 91 (2003) 057002.

[13] M. Naito, M. Hepp, Jpn. J. Appl. Phys. 39 (2000) L485.

[14] M. Naito, S. Karimoto, A. Tsukada, Supercond. Sci. Technol. 15 (2002) 1663.

[15] D. M. King, Z. –X. Shen, D. S. Dessau, B. O. Wells, W. E. Spicer, A. J. Arko, D. S. Marshall, J. DiCarlo, A. G. Loeser, C. H. Park, E. R. Ratner, J. L. Peng, Z. Y. Li, R. L. Greene, Phys. Rev. Lett. 70 (1993) 3159.

[16] N. P. Armitage, F. Ronning, D. H. Lu, C. Kim, A. Damascelli, K. M. Shen, D. L. Feng, H. Eisaki, Z. –X. Shen, Phys. Rev. Lett. 88 (2002) 257001.

[17] M. Naito, H. Sato, H. Yamamoto, Physica C 293 (1997) 36.

[18] A. Shibayama, T. Kitayama, T. Hayasaka, S. Ido, Y. Uno, T. Hosokawa, J. Nakata,

K. Nishimura, M. Nakajima, Rev. Sci. Instrum. 60 (1989) 1779.

[19] F. M. Quinn, D. Teehan, M. MacDonald, S. Downes, P. Bailey, J. Synchrotron Rad. 5 (1998) 783.

[20] P. A. P. Lindberg, Z. –X. Shen, W. E. Spicer, I. Lindau, Surf. Sci. Rep. 11 (1990) 1.

[21] T. Sato, T. Kamiyama, T. Takahashi, K. Kurahashi, K. Yamada, Science 291 (2001) 1517.

[22] D. S. Dessau, Z. –X. Shen, D. M. King, D. S. Marshall, L. W. Lambardo, P. H. Dickinson, A. G. Loeser, J. DiCarlo, C. –H. Park, A. Kapitulnik, W. E. Spicer, Phys. Rev. Lett. 71 (1993) 2781.

[23] S. Massidda, N. Hamada, J. Yu, A. J. Freeman, Physica C 157 (1989) 571.

[24] N. P. Armitage, D. H. Lu, C. Kim, A. Damascelli, K. M. Shen, F. Ronning, D. L. Feng, P. Bogdanov, Z. –X. Shen, Y. Onose, Y. Taguchi, Y. Tokura, P. K. Mang, N. Kaneko, M. Greven, Phys. Rev. Lett. 87 (2001) 147003.

**Figure captions**

Fig. 1    Energy distribution curve integrated approximately in the hatched region.

Fig. 2    Time evolution of the spectra: (a) as-transferred, (b) after 1 day, and (c) after 1 week.    Every EDC is integrated approximately in the hatched region.    Inset displays the near-$E_F$ region.

Fig. 3    Energy distribution curves from various directions in the Brillouin zone obtained at $h\nu = 50$ eV.    (a) $\Gamma - (\pi, \pi)$, (b) $(\pi, 0) - (\pi, \pi)$, (c) $(0, 0) - (\pi, 0)$, (d) a line depicted in the figure.

Fig. 4    Fermi surface of the partial Brillouin zone of LCCO ($x = 0.1$).    Circles, triangles, and squares denote the $E_F$ crossing points determined using EDCs from lines parallel to $\Gamma–(\pi, \pi)$, $(\pi, 0)–(\pi, \pi)$, $(\pi, 0)–(0, \pi)$ directions, respectively.    Circles and triangles were from the data taken over an approximate Brillouin zone octant and symmetrized across the $\Gamma$ to $(\pi, \pi)$ line, while squares were from those taken over an almost full quadrant.

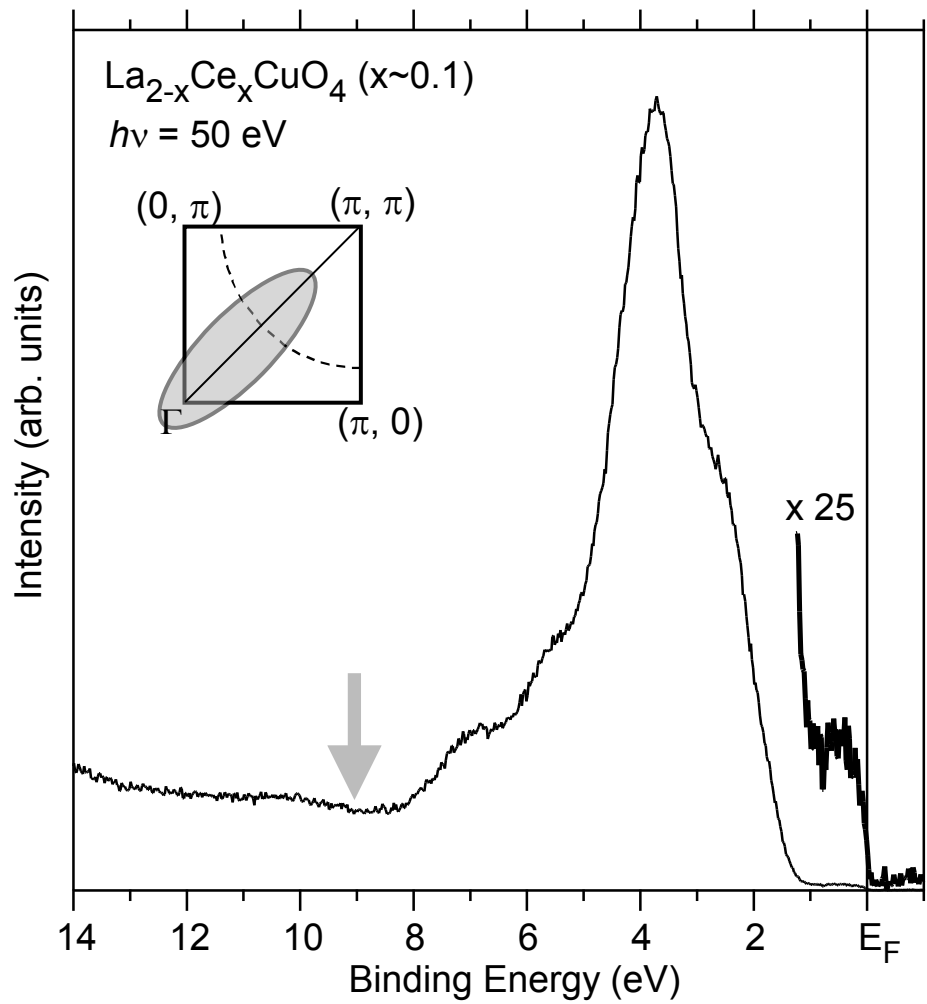

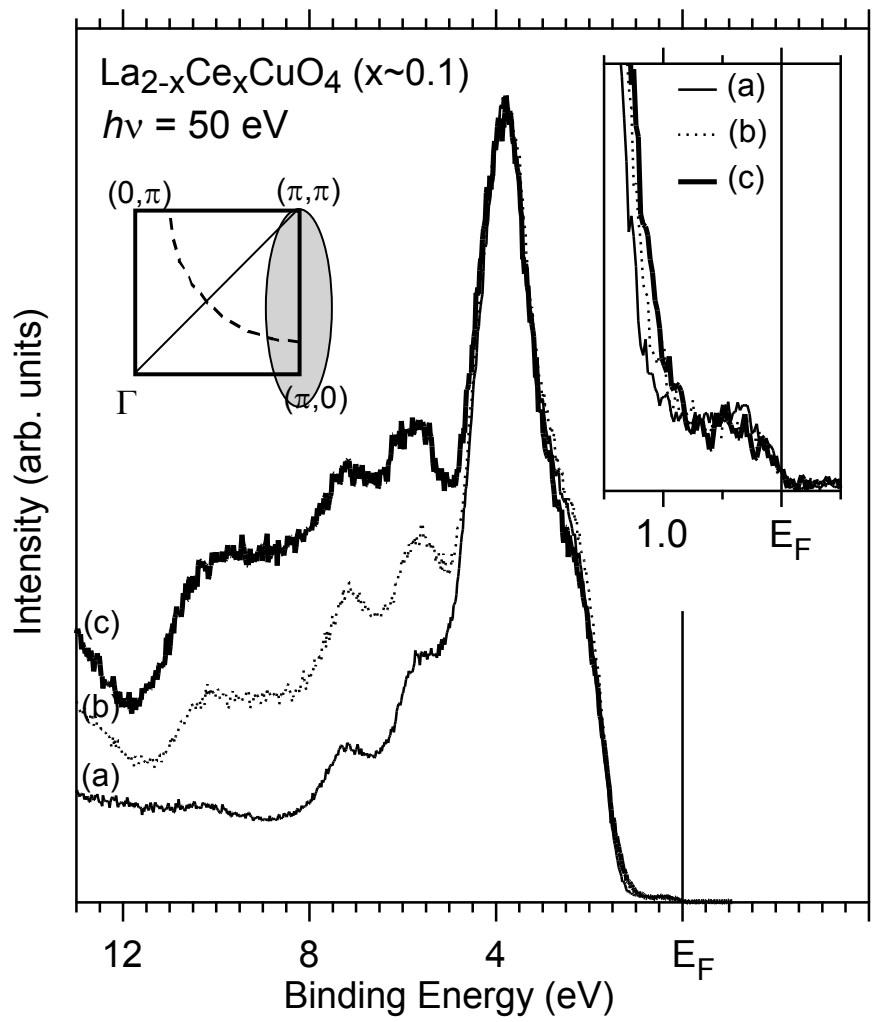

H. Yamamoto et al. Fig. 2
PC-24/ISS2003

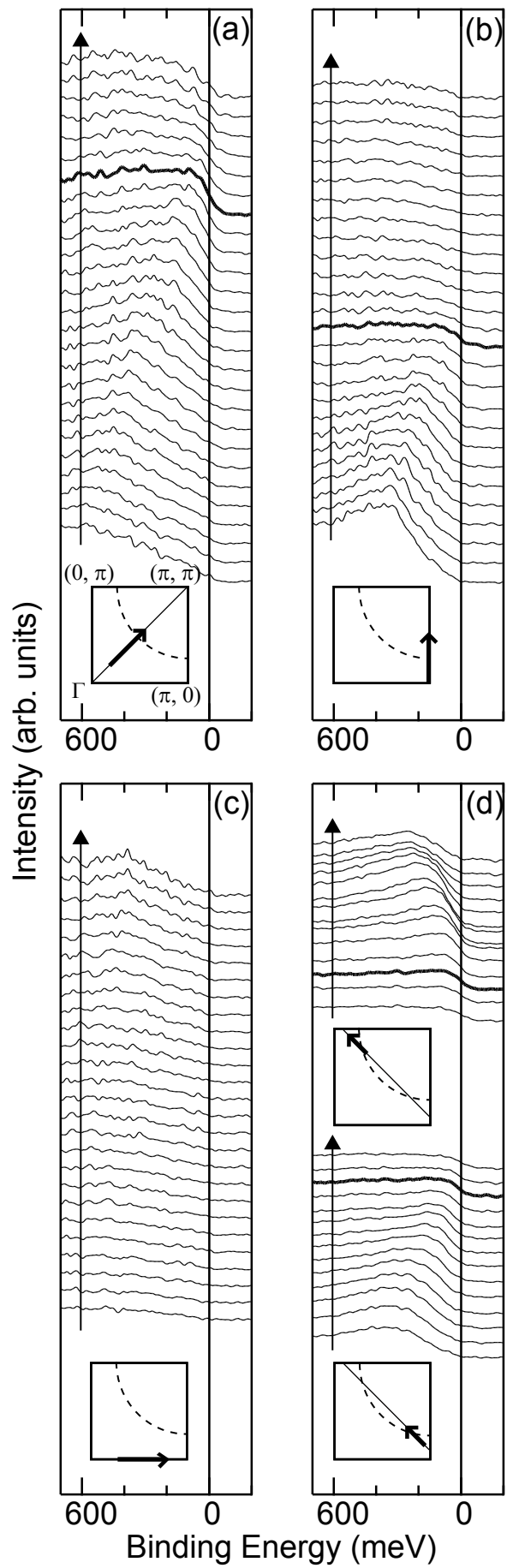

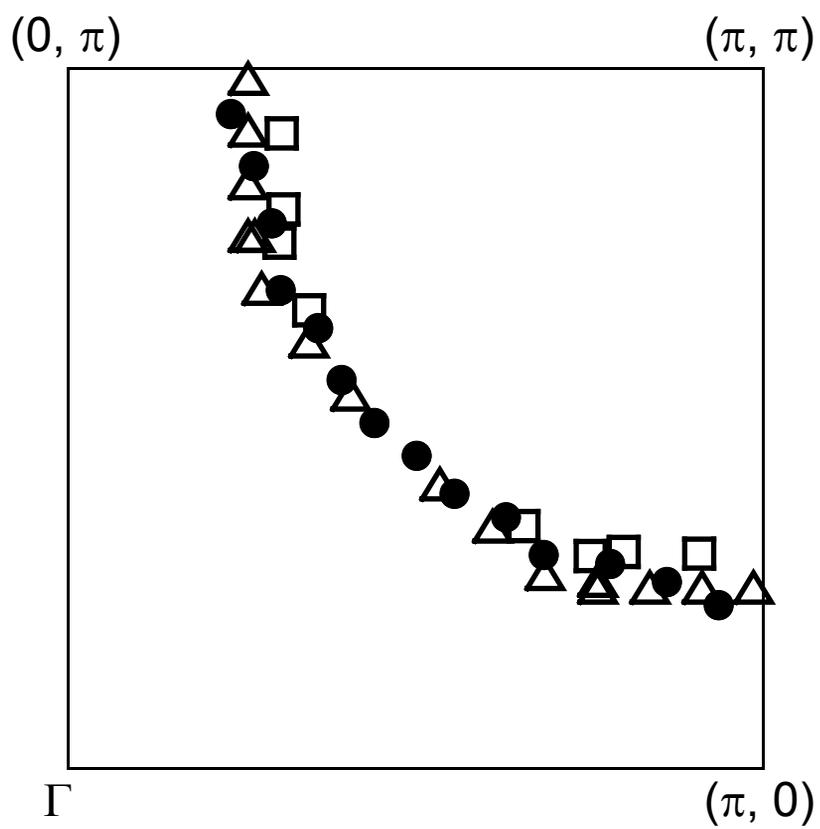